\documentclass{aastex63}

\usepackage{dcolumn}
\usepackage{bm}
\usepackage{xcolor}
\usepackage{multibib}
\newcites{main}{References}
\newcites{appendix}{References (Appendix)}
\usepackage{subfigure}
\usepackage{soul}
\usepackage{tcolorbox}
\usepackage{amsmath}
\usepackage{verbatim}

\newcommand{\beq}{\begin{equation}}
\newcommand{\eeq}{\end{equation}}
\newcommand{\bse}{\begin{subequations}}
\newcommand{\ese}{\end{subequations}}
\newcommand{\bary}{\begin{eqnarray}}
\newcommand{\eary}{\end{eqnarray}}
\newcommand{\bwt}{\begin{widetext}}
\newcommand{\ewt}{\end{widetext}}

\begin{document}

\title{Extreme HBL-like behavior of Markarian 421 and its two-zone photohadronic interpretation
}


\author{Sarira Sahu}
\email{sarira@nucleares.unam.mx}
\affiliation{Instituto de Ciencias Nucleares, Universidad Nacional Aut\'onoma de M\'exico, \\
Circuito Exterior, C.U., A. Postal 70-543, 04510 Mexico DF, Mexico}

\author{Carlos E. L\'opez Fort\'in}
\email{carlos.fortin@correo.nucleares.unam.mx}
\affiliation{Instituto de Ciencias Nucleares, Universidad Nacional Aut\'onoma de M\'exico, \\
Circuito Exterior, C.U., A. Postal 70-543, 04510 Mexico DF, Mexico}

\author{Isabel Abigail Valadez Polanco}
\email{abivaladez@gmail.com}
\affiliation{Facultad de Ingeniería, Universidad Aut\'onoma de Yucat\'an, \\
Industrias No Contaminantes S/N, Sin Nombre de Col 27, M\'erida, Yucat\'an, M\'exico}



\author{Subhash Rajpoot}
\email{Subhash.Rajpoot@csulb.edu}
\affiliation{Department of Physics and Astronomy, California State University,\\ 
1250 Bellflower Boulevard, Long Beach, CA 90840, USA}



\begin{abstract}
Markarian 421 is the nearest high-energy peaked blazar and also  is
the first extragalactic source to be detected in multi-TeV
$\gamma$-rays. It has been observed in multiwavelength for
exceptionally long period of time with dense monitoring and several
major outbursts have been detected from this source. In March 2010,
the source was in a high state of activity and was observed in
multiwavelength by various telescopes for 13 consecutive days. During
this period the position of the synchrotron peak was found to be above $10^{17}$ Hz and also the position of the second peak was shifted towards higher energy, a signature of extreme HBL-like behavior. We observed that the standard photohadronic model is inadequate to explain the observed spectra. However, a recently proposed two-zone photohadronic model explains very well the GeV-TeV flaring events observed by both MAGIC and VERITAS telescopes. From the observation of the highest energy $\gamma$-ray event on MJD 55266 we also estimated the minimum bulk Lorentz factor.
\end{abstract}

\keywords{High energy astrophysics (739), Blazars (164), Gamma-rays (637), Relativistic jets (1390), BL Lacertae objects (158)}

\section{Introduction}

The spectral energy distributions (SEDs) of  blazars are characterized by two non-thermal peaks \citep{Dermer:1993cz}. The first peak, in the infrared to X-ray energy range, is produced by the synchrotron emission from  relativistic electrons in the jet. The second peak, in between X-ray to $\gamma$-ray energy, is believed to be produced either from the synchrotron Self-Compton (SSC) scattering of high-energy electrons with the low-energy self-produced synchrotron photons in the jet \citep{Maraschi:1992iz,Murase:2011cy,Gao:2012sq} or from the external Compton (EC) scattering  with external sources such as photons from the accretion disk or broad-line regions \citep{Sikora:1994zb,Blazejowski:2000ck}. The model which explains these two peaks is in general referred to as the leptonic model. Depending on the position of the synchrotron peak, the blazars are classified into four categories:  low-energy peaked blazars (LBLs, $\nu_{peak} < 10^{14}\, Hz$), intermediate-energy peaked blazars (IBLs, $\nu_{peak}$ between $10^{14}\, Hz$ and $10^{15}\, Hz$), high energy-peaked blazars (HBLs, $\nu_{peak}$ between $10^{15}\, Hz$ and $10^{17}\, Hz$) \citep{Abdo:2009iq} and extreme energy peaked blazar (EHBL, $ >10^{17}$ Hz) \citep{Costamante:2001pu}. In particular, the  EHBLs are characterized by low luminosity and have limited variabilities which make these objects difficult to detect \citep{Costamante:2017xqg,Acciari:2019jzj}.

So far, only a few EHBLs have been detected in the very high energy (VHE, $>\, 100$ GeV) regime and among them some well known objects are 1ES 0229+200, 1ES 0347-232, RGB J0710+591 and 1ES 1101-232 \citep{Costamante:2017xqg}. Among all the EHBLs observed so far, the highest energy peak frequencies correspond to 1ES 0229+200 with the synchrotron peak at $\nu^{peak}_{syn}\simeq 3.5\times 10^{19}$ Hz and the SSC peak at $\nu^{peak}_{IC}\simeq 1.5\times 10^{27}$ Hz \citep{Kaufmann:2011az} respectively.
The nearest and the extensively studied HBLs Markarian 421 (Mrk 421), Markarian 501 (Mrk 501) and 1ES 1959+650 have also shown EHBL-like behavior with harder TeV spectra in which the synchrotron and SSC peaks are shifted towards higher energy during some flaring episodes \citep{Ahnen:2018mtr,Hayashida:2020wez,Aleksi__2015}. The EHBL-like behavior of Mrk 501 was observed during two episodes of very-high energy flaring events of May - July 2005 and June 2012. Also, during a multiwavelength (MW) campaign of 1ES 1959+650 between 29th April to 21st November 2016, the MAGIC telescopes observed multi-TeV flaring events during the nights of 13th, 14th June and 1st July 2016 when the position of the synchrotron peak was in EHBL range and exhibiting extreme HBL-like behavior. 

It is very difficult to explain the shift of the second peak in the one-zone leptonic model. However, it has been shown that by using unrealistic model parameters, such as the large values of minimum electron Lorentz factor and the bulk Lorentz factor \citep{Ahnen:2018mtr,Hayashida:2020wez} as well as a low magnetic field, the shift can be explained. Alternative solutions have also been proposed, such as: a two-zone leptonic model, IC scattering of the electron with the cosmic microwave background, a spine-layer structured jet model and variants of the hadronic model \citep{Boettcher:2008fh,Acciari:2019ntl,Ahnen:2018mtr}.

In previous studies, \cite{Sahu:2019kfd,Sahu:2019scf} have used the photohadronic model to explain the multi-TeV flaring from many HBLs very well. 
In all these studies it is shown that the Fermi accelerated high energy protons 
interacting with the seed SSC photons in the jet can explain the VHE spectrum very 
well. The SSC photon flux $\Phi_{SSC}$ in this region is a perfect power-law given as
$\Phi_{SSC}\propto \epsilon_{\gamma}^{\beta} \propto E_{\gamma}^{-\beta}$ (in units of $\mathrm{erg}\ \mathrm{cm^{-2}}\ \mathrm{s^{-1}}$), where
$\epsilon_{\gamma}$ is the seed photon energy, $E_{\gamma}$ is the energy of the
observed VHE gamma-ray and $\beta$ is the seed photon spectral index. The value of $\beta$ lies in the range $0 < \beta \le 1$.
However, recently, \cite{Sahu:2020tko,Sahu:2020kce} have shown that this standard
photohadronic scenario is inadequate to explain the EHBL-like behavior of the flaring
events of Mrk 501 and 1ES 1959+650. It was also shown that the seed SSC photon flux in
the kinematically allowed region for the $\Delta$-resonance production from $p\gamma$
interaction is no more a single power-law due to the EHBL-like behavior of the flaring
events. So, the standard photohadronic  model was extended to include two different zones for the SSC flux. The zone-1 (low energy region) has the conventional behavior having $0 < \beta \le 1$. But in zone-2 (high energy region) the value of $\beta$ is different and in the range   
$ 1 < \beta  \le 1.5$. However, the spectral index $\alpha$ of Fermi accelerated high energy protons in both the zones remain the same. 

During the EHBL-like flaring events of Mrk 501 and 1ES 1959+650, the positions of the synchrotron peak and the SSC peak were shifted towards the higher energies. So, \cite{Sahu:2020tko,Sahu:2020kce} assumed that the transition from HBL to EHBL-like behavior is due to the shift of the synchrotron peak and the SSC peak in the SED towards higher energies.
As a result of this shift, jet parameters such as bulk Lorentz factor, blob size, and
magnetic field may be different from HBL, but the particle acceleration and emission
mechanisms remain the same. Therefore, the photohadronic process should still be the
dominant process in this changed environment. It is also argued that these EHBL-like
events are transient in nature and may repeat in the future.

During a MW campaign in 2010, flaring in GeV-TeV energy was observed in Mrk 421. The VHE activity started decreasing from a high flux
to a normal flux during the period from March 10 to March 22, i.e. for 13 consecutive days \citep{Aleksi__2015}. It was 
observed that the positions of both the synchrotron peak and the SSC peak of Mrk 421 were shifted towards higher energy values consistent with the EHBL classification of the object, were similar to the
flaring of Mrk 501 and 1ES1959+650 discussed above. This shift in the peak positions towards higher energies is explained using electrostatic acceleration mechanism \citep{Zheng:2020nsk}. Also
recently it has been shown that these VHE spectra of Mrk 421 can be explained better in the
photohadronic model than the one and the two-zone leptonic models \citep{deLeon:2021lof}.  As
the two-zone photohadronic model has successfully explained the EHBL-like behavior of
Mrk 501 and 1ES1959+650, we use this model once again to explain the
EHBL-like behavior of Mrk 421 observed during March 2010.

\section{Flaring of Mrk 421}

Mrk 421 is the nearest ($z=0.031$) well known HBL with a central black hole mass 
$M_H\approx (2-9)\times 10^8 M_{\odot}$ and is the first extragalactic source to be
detected in multi-TeV $\gamma$-rays \citep{Punch:1992xw}. It is also one of the fastest
varying $\gamma$-ray sources. Since its discovery, the object has been studied
intensively through dedicated MW observations and several major flares have been
observed \citep{Amenomori:2003gj,Fossati:2007sj,Cui:2004wi,Abeysekara:2016qwu,Mastichiadis:2013aga,Beck:2020gdi}. Its average MW SED has been modelled with both leptonic \citep{Abdo:2011zz,Banerjee:2019oyp} and hadronic \citep{Cerruti:2014iwa,Zech:2017lma} models. During April-May 1994, a TeV/X-ray flare was reported and on May 14/15 a correlation between the TeV emission and X-ray flare was observed. Also, during a
multiwavelength campaign in 2004, large flares in X-ray by Rossi X-ray Timing Explorer (RXTE) and TeV flare by  Whipple 10 m telescope were observed \citep{Blazejowski:2005ih}. It is important to note
that during this period, the TeV flare had no coincident low energy counterparts, and
most significantly, 
the X-ray flux reached its peak 1.5 days before the TeV flux during this outburst.
A remarkable similarity between this VHE flare and the orphan TeV flare in 1ES 1959+650
of 2002 \citep{Krawczynski:2003fq} and a similar variation pattern in their X-ray
emission was established. Exceptionally long and dense monitoring in MW of Mrk 421 has
been undertaken since 2009 to understand the temporal evolution of the SED. Mrk 421 was
in active state during a MW campaign in 2010 and an extraordinary flare $\sim 27$ Crab
Unit (c.u.) above 1 TeV was first detected by VERITAS telescopes on February 16 and a
follow-up observation was undertaken by the HESS telescopes \citep{Tluczykont:2011gs}
for four subsequent nights. So far, this is the  highest flare ever observed from Mrk
421. Also, flaring in VHE gamma-ray was observed for 13 consecutive days from March 10
(MJD 55265) to March 22 (MJD55277). Again, during a 6-month long multi-instrument
campaign \citep{Banerjee:2019oyp},
a large VHE flare about 16 times larger than the usual one, was observed by MAGIC
telescopes on 25th of April 2014 and soon it was followed-up by XMM-Newton and VERITAS.
Several of the above multi-TeV flaring events from Mrk 421 have been explained extremely well by the standard photohadronic scenario \citep{Sahu:2015tua,Sahu:2018gik,Sahu:2019kfd}.

\section{Two-zone photohadronic model}
The photohadronic model is very successful in explaining the multi-TeV flaring events from many HBLs
\citep{Sahu:2019lwj,Sahu:2019kfd}. The epoch of VHE flaring is explained by assuming
the formation of a double jet structure along the same axis: an inner jet of size $R'_f$
and photon density $n'_{\gamma,f}$ 
buried under an outer jet of size $R'_b$ and
photon density $n'_{\gamma}$ where $R'_f<R'_b$ and $n'_{\gamma,f}>n'_{\gamma}$. Here
the notation $'$ implies comoving frame. The internal and the external jets are moving
with almost the same bulk Lorentz factor $\Gamma_{in}\simeq \Gamma_{ext}\simeq \Gamma$.
Also, their common Doppler factor is $\mathcal{D}$ (for blazars $\Gamma \simeq
\mathcal{D}$). The geometrical structure of the jet and the detailed description of the photohadronic model is given in Fig. 1 of \cite{Sahu:2019lwj}. 

In the inner jet region the Fermi accelerated protons with a differential power-law spectrum
$dN/dE_p\propto E^{-\alpha}_p$ ($\alpha \ge 2$) 
interact with the background SSC seed
photons to produce the $\Delta$-resonance and its subsequent decay gives $\gamma$-rays and neutrinos via the decay of neutral pion and charged pion respectively. As the inner region is hidden, there is no way to directly estimate the photon
density there. Thus we assume a simple scaling behavior of the photon densities in the inner and the outer jet regions satisfying the criteria
%
\beq
\frac{n'_{\gamma,f}(\epsilon_{\gamma,1})}{n'_{\gamma,f}(\epsilon_{\gamma,2})} \simeq\frac{n'_{\gamma}(\epsilon_{\gamma,1})}{n'_{\gamma}(\epsilon_{\gamma,2})}.
\label{eq:scal}
\eeq
%
The above equation implies that the ratio of photon densities at two different background energies $\epsilon_{\gamma,1}$ and $\epsilon_{\gamma,2}$ in the inner and the outer jet regions are almost the same. By using the relation in
Eq. (\ref{eq:scal}), we can express the inner photon density in terms of the observed flux. The kinematical condition \citep{Sahu:2019lwj} to produce the $\Delta$-resonance is, 
%
\beq
E_p \epsilon_\gamma=\frac{0.32\ \Gamma{\mathcal D}}{(1+z)^{2}}\ \mathrm{GeV^2},
\label{eq:kinproton}
\eeq
%
where $z$ is the redshift of the object. The observed VHE $\gamma$-ray energy
$E_{\gamma}$ and the proton energy $E_p$ are related through $E_{\gamma}\simeq
0.1\,E_p$. In previous works by \cite{Sahu:2019lwj} it is explicitly shown that for
the production of $\Delta$-resonance, the value of $\epsilon_{\gamma}$ always lies in
the lower tail region of the SSC band and for HBLs the flux in this region is a perfect power-law.

The VHE $\gamma$-rays en route to Earth undergo
energy dependent attenuation  by the extragalactic background light (EBL)
through electron-positron pair production and this pair-production process
not only attenuates the absolute flux but also significantly changes the 
shape of the VHE spectra. The measurement of the EBL is very 
difficult due to the uncertainties in the contribution of
zodiacal light \citep{Hauser:2001xs,Chary:2010dc} and consequently
galaxy counts result in a lower limit since the number of unresolved
sources are unknown \citep{Pozzetti:2000ew}. Several approaches with different
degrees of complexity have been developed to
calculate the EBL density as a function of energy for different
redshifts \citep{Dominguez:2010bv,Inoue:2012bk,Franceschini:2008tp,Stecker:2005qs,Stecker:2016fsg}. For the present analysis we use the EBL model of  \cite{Franceschini:2008tp}, which is consistent with a EBL peak flux of $\sim 15\ \mathrm{nW}\mathrm{n^{-2}}\ \mathrm{sr^{-1}}$ at $1.4\, \mathrm{\mu m}$ and is also widely used in other leptonic and hybrid models \citep{Aleksic:2014usa,Ahnen:2016hsf}. Taking into account the EBL contribution the observed
VHE flux from the sources can be given as \citep{Sahu:2019lwj}
%
\beq
F_{\gamma,obs}(E_{\gamma})=A_{\gamma} \Phi_{SSC}(\epsilon_{\gamma}) \left ( \frac{E_\gamma}{TeV} \right )^{-\alpha+3}\,e^{-\tau_{\gamma\gamma}(E_\gamma,z)},
\label{eq:fluxphi}
\eeq
%
where $A_{\gamma}$ is a dimensionless constant and $\tau_{\gamma\gamma}$ is the optical depth for the $\gamma\gamma\rightarrow e^+e^-$
process which depends on $E_{\gamma}$ and the redshift.
In the standard photohadronic model, $\Phi_{SSC}\propto E^{-\beta}_{\gamma}$ and putting this in 
Eq. (\ref{eq:fluxphi}) the flux can be written as

%
\beq
F_{\gamma,obs}(E_{\gamma})=F_0 \left ( \frac{E_\gamma}{TeV} \right )^{-\delta+3}\,e^{-\tau_{\gamma\gamma}(E_\gamma,z)}=F_{\gamma, in}(E_{\gamma})\, e^{-\tau_{\gamma\gamma}(E_\gamma,z)}.
\label{eq:fluxgeneral}
\eeq
%
The normalization constant $F_0$ is deduced from the observed VHE spectrum.
Thus the  spectral index $\delta=\alpha+\beta$ is the only free parameter in this model and
$F_{\gamma,in}$ is the intrinsic VHE $\gamma$-ray flux. 
In previous studies \citep{Sahu:2020tko,Sahu:2020kce} it was
shown that the EHBL-like behavior of the flaring epoch can not be explained by
the canonical photohadronic model. In order to remedy this defect the tail region of the SSC band which is
responsible for the production of the $\Delta$-resonance is assumed to have two different
power-laws with spectral indices $\beta_1$ (lower part of the tail region) and $\beta_2$ (upper part of the tail region).
Thus the photon flux is expressed as  
%
\beq
\Phi_{SSC}\propto
 \left\{ 
\begin{array}{cr}
E^{-\beta_1}_{\gamma}
, & \quad 
 \mathrm{100\, GeV\, \lesssim E_{\gamma} \lesssim E^{intd}_{\gamma}}
\\ E^{-\beta_2}_{\gamma} ,
& \quad   \mathrm{E_{\gamma}\gtrsim E^{intd}_{\gamma}}
\\
\end{array} \right. .
\label{eq:sscflux}
\eeq
%
Here $\beta_1\neq \beta_2$ and $E^{intd}_{\gamma}$ is an energy scale around which the transition takes place and its value can be fixed from the individual flaring spectrum. 
We suppose that the change in the value of the spectral index from $\beta_1$ to  $\beta_2$ around $E^{intd}_{\gamma}$ could be due to the effect of magnetic field on the energy loss process of the high energy electrons to the SSC photons or it could be some new physics. However, we do not dwell on the transition between these two zones as the exact physical mechanism behind it is unclear at this stage. The value of $E^{intd}_{\gamma}$ corresponds to a value of $\epsilon_{\gamma}$ in the SSC band. This separates $\Phi_{SSC}$ into two distinct regions. By inserting Eq. (\ref{eq:sscflux}) into Eq. (\ref{eq:fluxphi}), the observed VHE spectrum can be expressed as
\beq
F_{\gamma, obs}=
e^{-\tau_{\gamma\gamma}}\times
\begin{cases}
 F_1 \, \left ( \frac{E_{\gamma}}{TeV} \right )^{-\delta_1+3}
, & \quad 
\mathrm{100\, GeV\, \lesssim E_{\gamma} \lesssim E^{intd}_{\gamma}}\,\,\,\,\, (\text{zone-1}) \\ 
F_2 \, \left ( \frac{E_{\gamma}}{TeV} \right )^{-\delta_2+3},
& \quad \,\,\,\,\,\,\,\,\,\,\,\,\,\,\,\,\,\,\,\,\,\,\,\,\,\,\,\,\,\,\, \mathrm{E_{\gamma}\gtrsim E^{intd}_{\gamma}}\,\,\,\,\, (\text{zone-2})
 \end{cases}.
\label{eq:flux}
\eeq
%
In this two-zone photohadronic model $F_1$ and $F_2$ are the normalization constants and the spectral indices $\delta_i=\alpha+\beta_i$ ($i=1,2$) are the free parameters to be adjusted by fitting the model to the observed VHE spectrum of the EHBL. The proton spectral index is $\alpha\ge 2.0$, and we use the generally accepted value of $\alpha=2.0$ \citep{Dermer:1993cz}. This automatically constrains the value of $\beta_i$ for a given $\delta_i$. However, it is important to mention that the value of $\delta_1$ (zone-1) allows one to characterize the flaring state for a given observation without depending explicitly on the modeling of the SSC region. In other words, in the photohadronic scenario, the low energy region of the VHE spectrum decides the flaring state of the outburst as the value of $\delta_1$ always lies in the range $2.5\le \ \delta \le 3.0$ \citep{Sahu:2019kfd}.

To fit the observed data for each VHE spectrum, we use a standard optimization of the parameter space through the mean squared error (MSE) with respect to the central values of the data points, and choose the best pair of ($F_{0}, \delta$) by maximizing the Pearson chi-squared ($\chi^2$) statistic and calculating the corresponding statistical significance.

The observed VHE spectrum of each day is fitted in the following manner. For the low energy regime of the spectrum we take $2.5 \le \delta_1 \le 3.0$ and fit it with different values of $F_1$. Similarly in the high energy regime  we take $3.1 \le \delta_2 \le 3.5$ and fit the spectrum for different values of $F_2$. From both the zones we select the best fits by calculating their statistical significance. The crossing point of these two curves gives the value of  $E^{intd}_{\gamma}$. Also, we have to make sure that these two curves intersect only once.

\begin{figure}
\begin{subfigure}
\centering
\includegraphics[width=.5\linewidth]{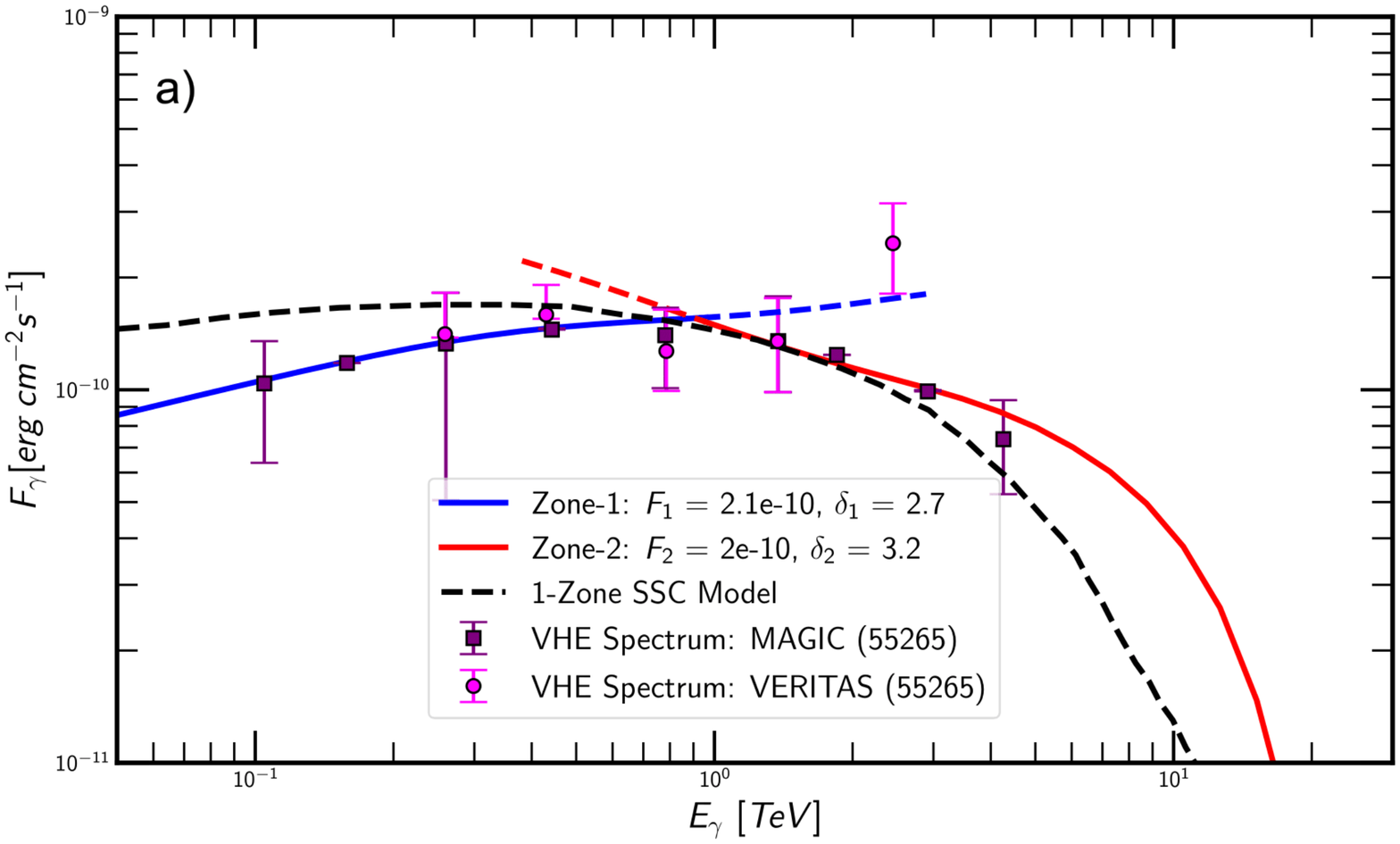}
\end{subfigure}
\begin{subfigure}
  \centering
  \includegraphics[width=.5\linewidth]{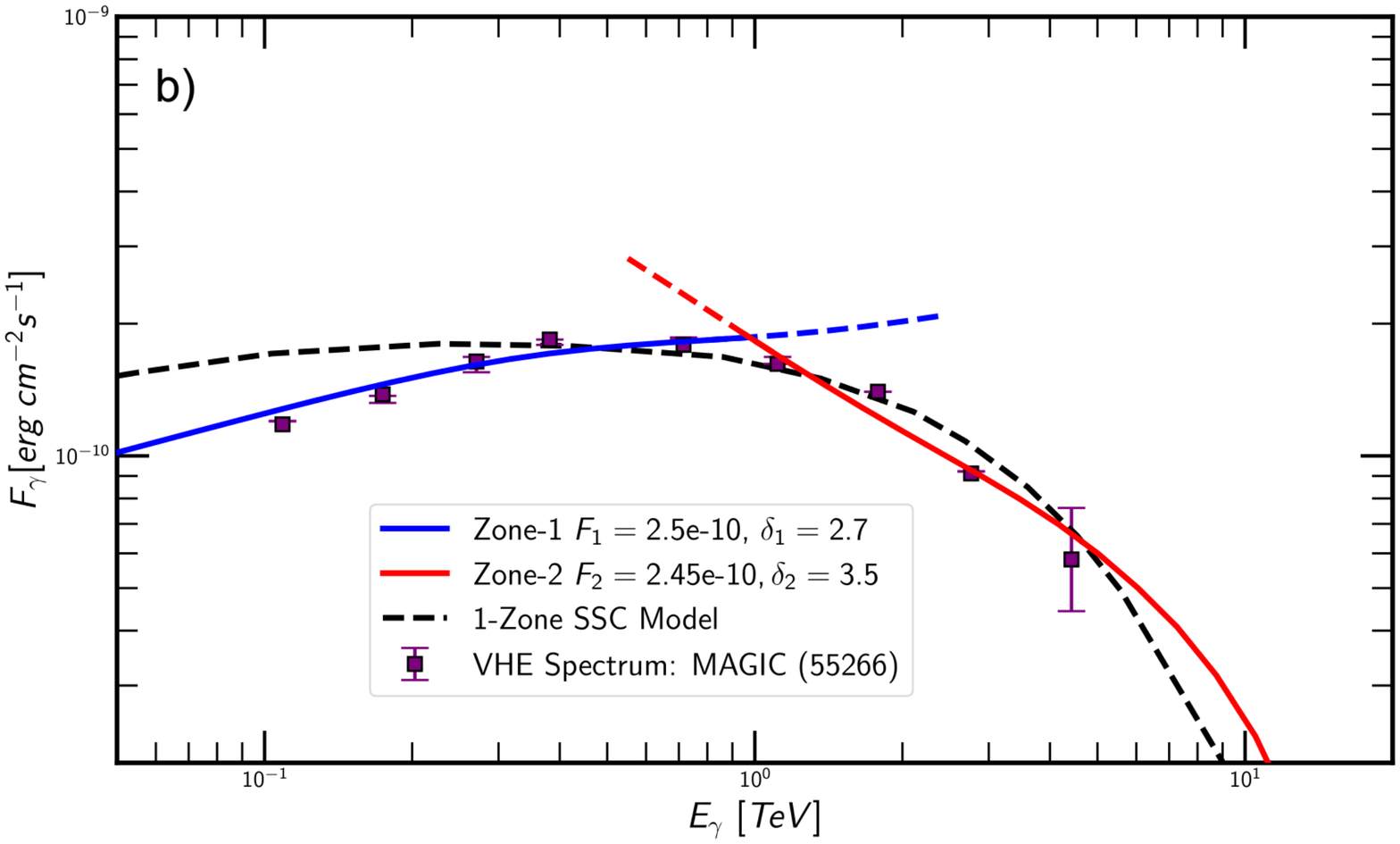}
\end{subfigure}%
\caption{The VHE flaring of MJD 55265 observed by MAGIC and VERITAS telescopes are shown in (a). Also the VHE event of MJD 55266 observed by MAGIC is shown in (b). Along with the two-zone photohadronic fit (blue for zone-1 and red for zone-2) the one-zone SSC fit for both days are shown for comparison.
In all the figures the normalization constants $F_i$ ($i=1,2$) are defined in units of $\mathrm{erg\, cm^{-2} \, s^{-1}}$. The fitted photohadronic model curves in zone-1 and zone-2 are extended by dashed line of their respective color beyond the transition energy just to show their likely behavior.}
\label{fig:figure1}
\end{figure}
\begin{figure}
\begin{subfigure}
\centering
  \includegraphics[width=.5\linewidth]{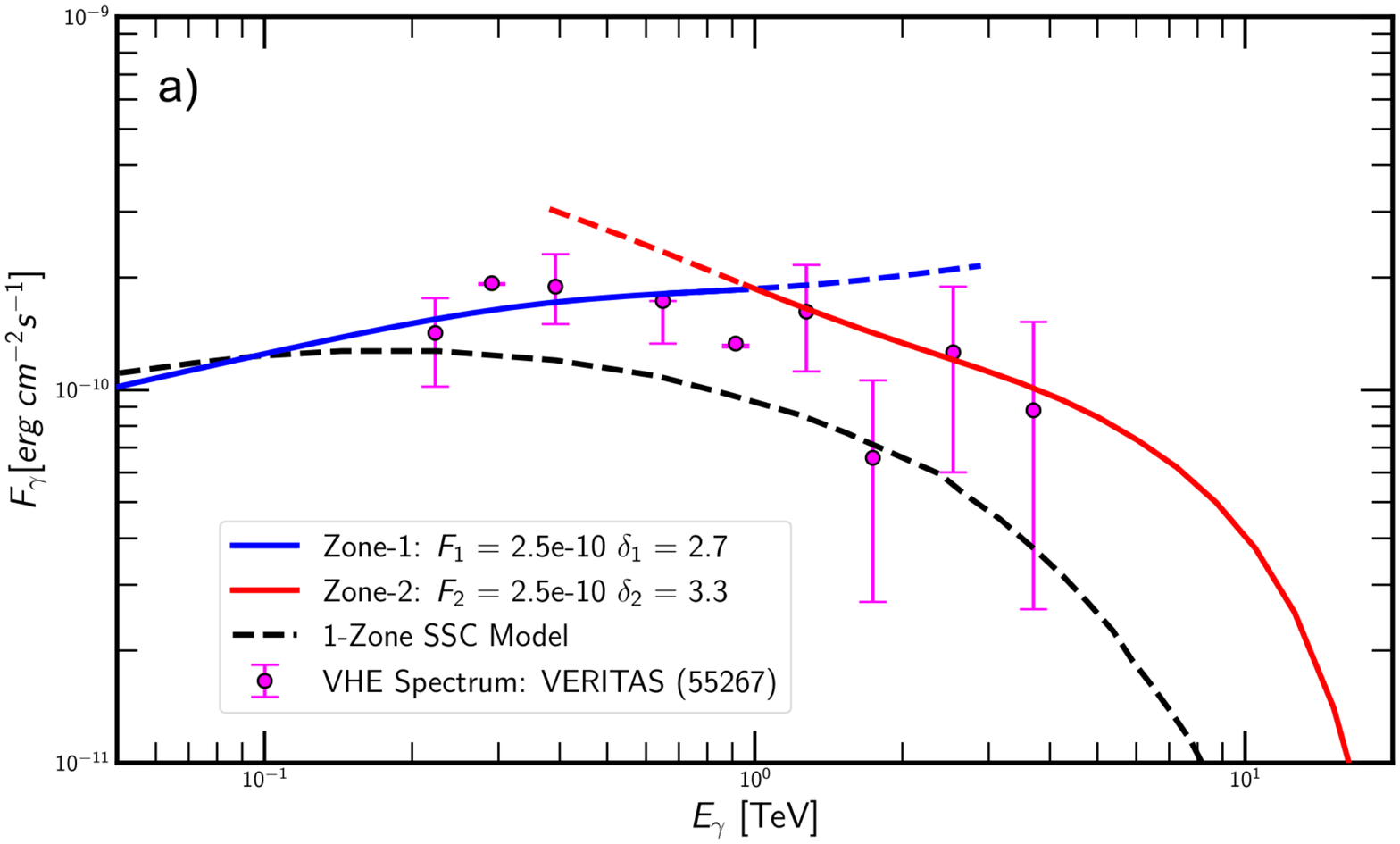}
\end{subfigure}
\begin{subfigure}
\centering
  \includegraphics[width=.5\linewidth]{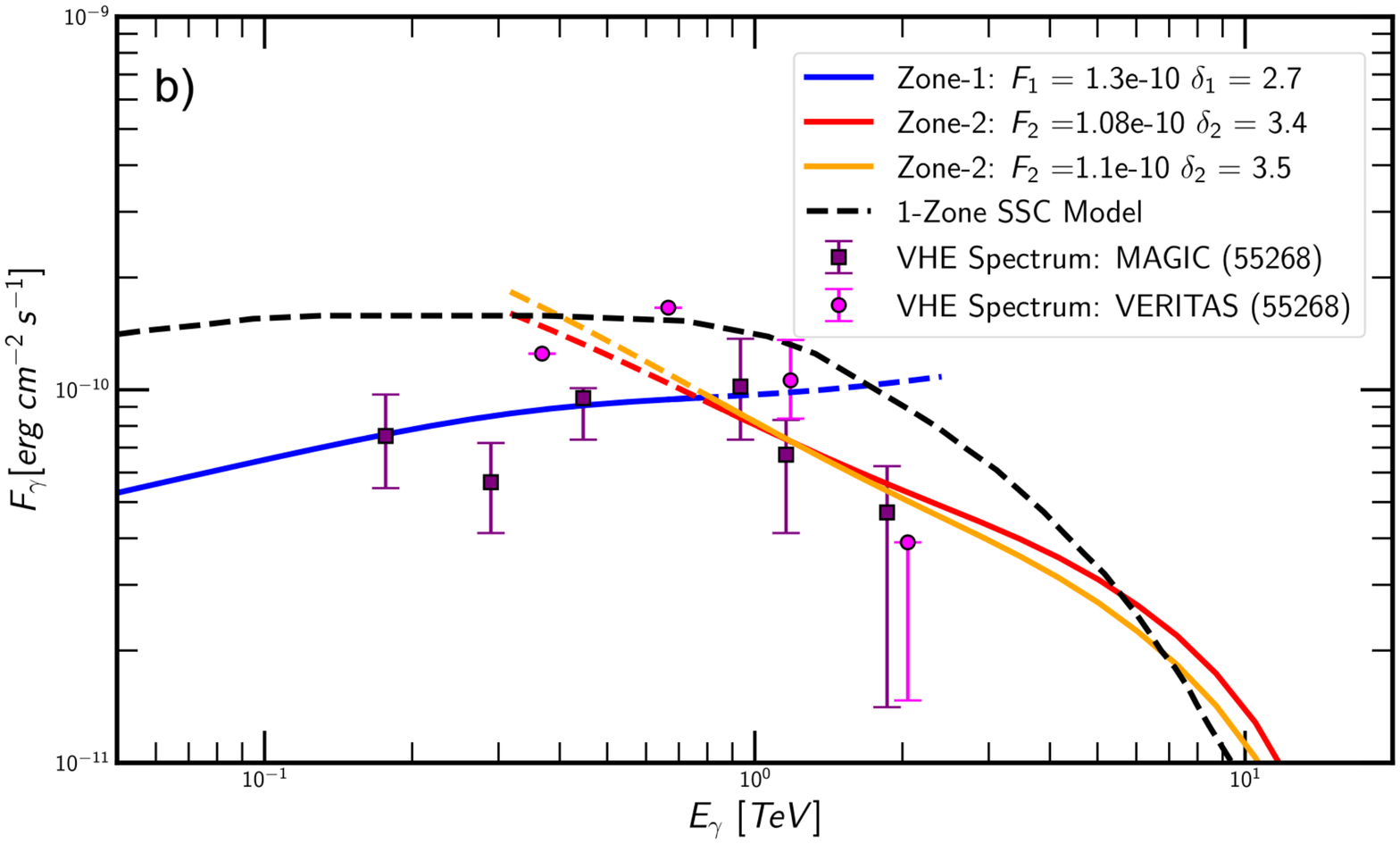}
\end{subfigure}%
\caption{The VHE flaring data of MJD 55267 observed by VERITAS (a) and of MJD 55268 by both MAGIC and VERITAS (b) are fitted with two-zone photohadronc mdoel which are shown along with the one-zone SSC fit for comparison.}
\label{fig:figure2}
\end{figure}
\begin{figure}
\begin{subfigure}
\centering
  \includegraphics[width=.5\linewidth]{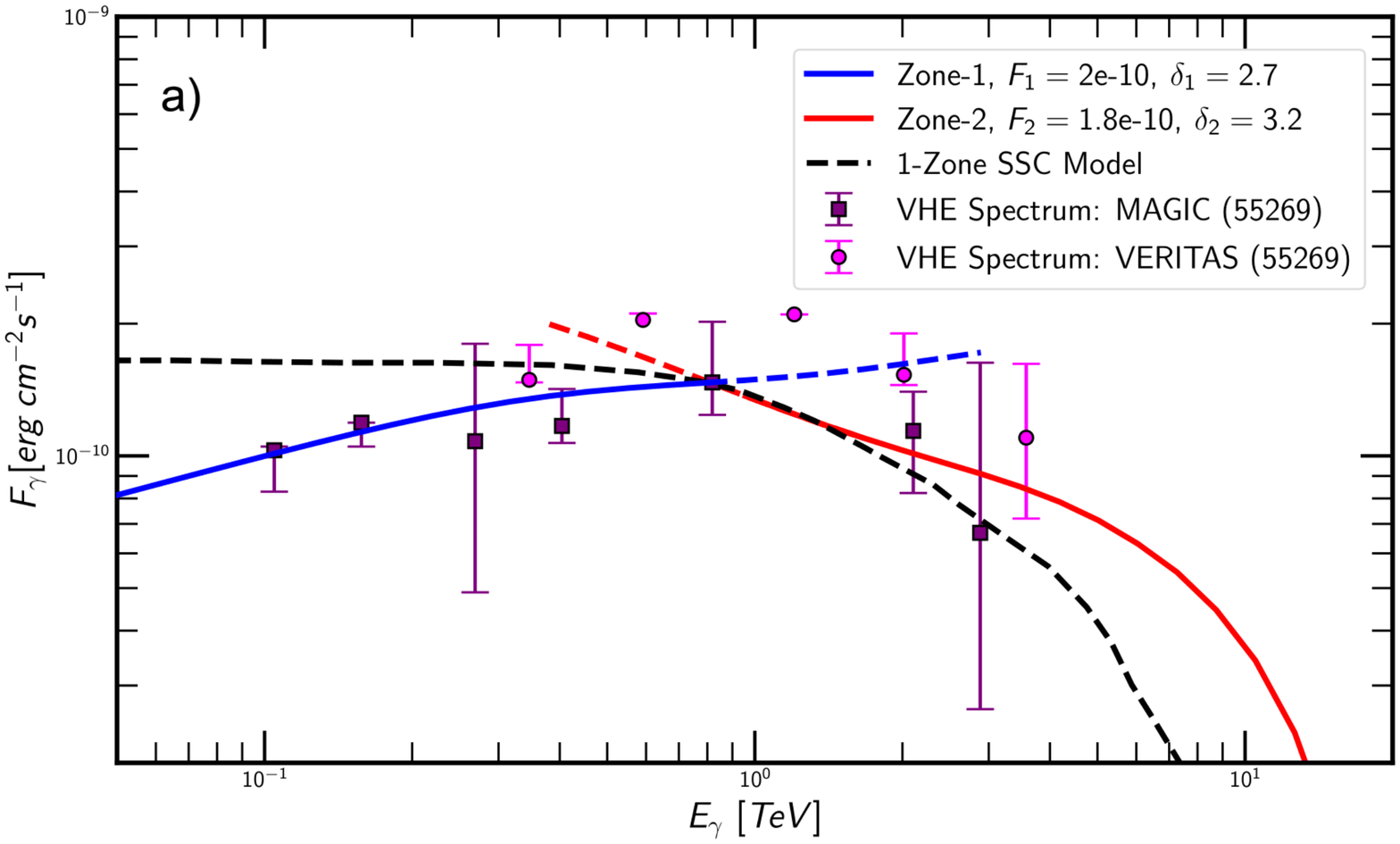}
\end{subfigure}
\begin{subfigure}
\centering
  \includegraphics[width=.5\linewidth]{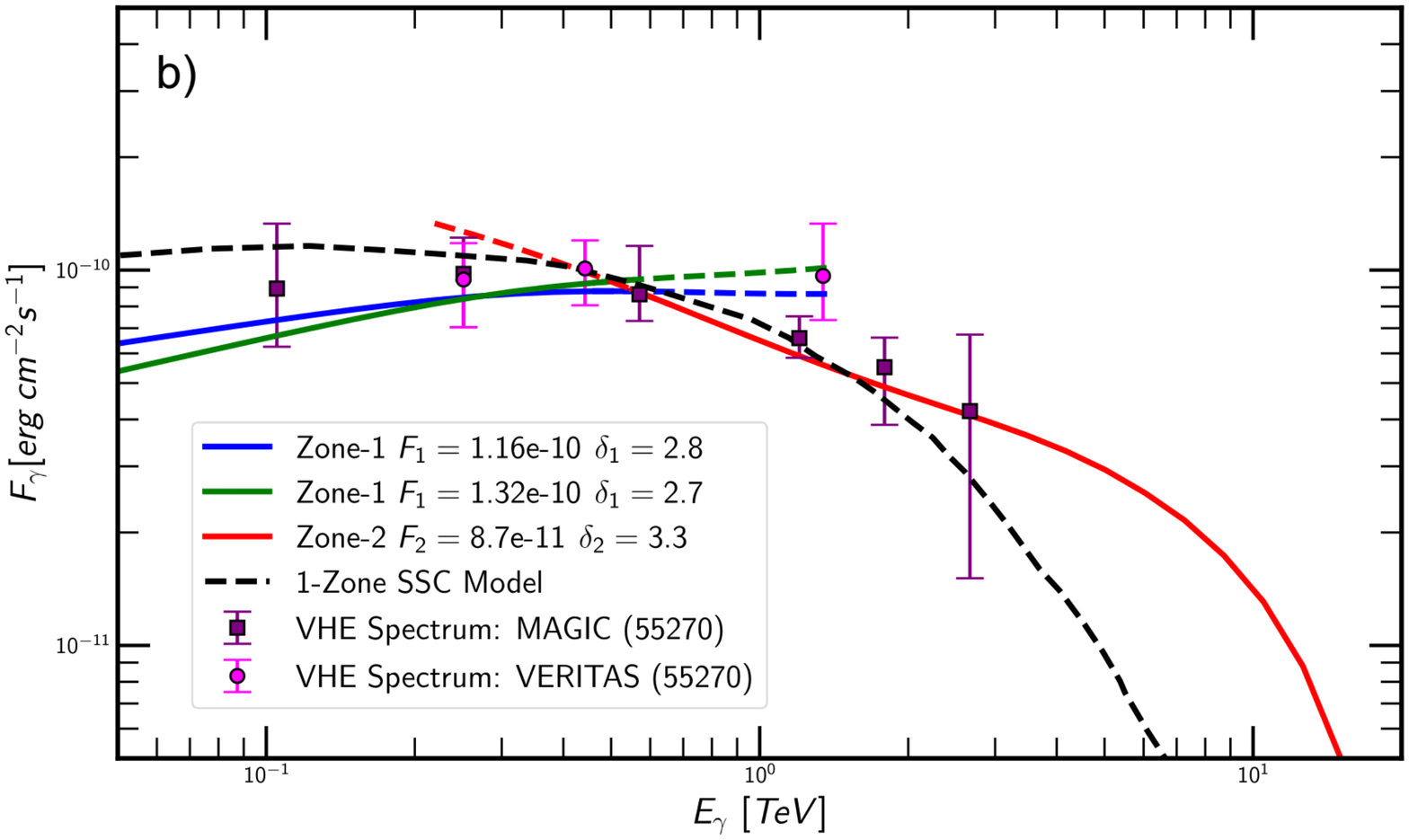}
\end{subfigure}%
\caption{The VHE spectra of MJD 55269 (a) and of MJD 55270 (b) observed by both MAGIC and VERITAS telescopes are fitted with two-zone photohadronic model. We have also shown the one-zone SSC fit for comparison.}
\label{fig:figure3}
\end{figure}
\begin{figure}
\begin{subfigure}
\centering
  \includegraphics[width=.5\linewidth]{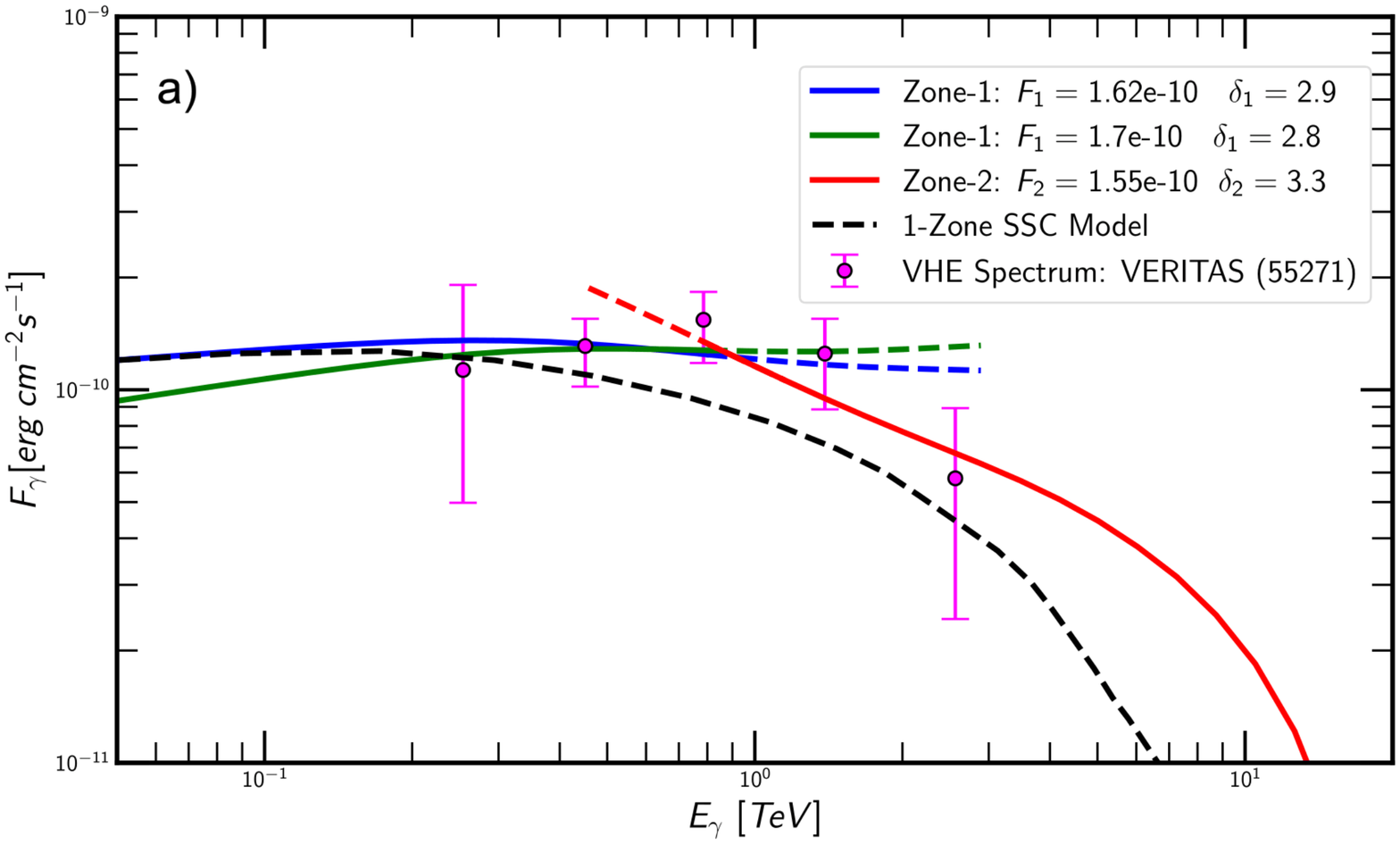}
\end{subfigure}
\begin{subfigure}
\centering
  \includegraphics[width=.5\linewidth]{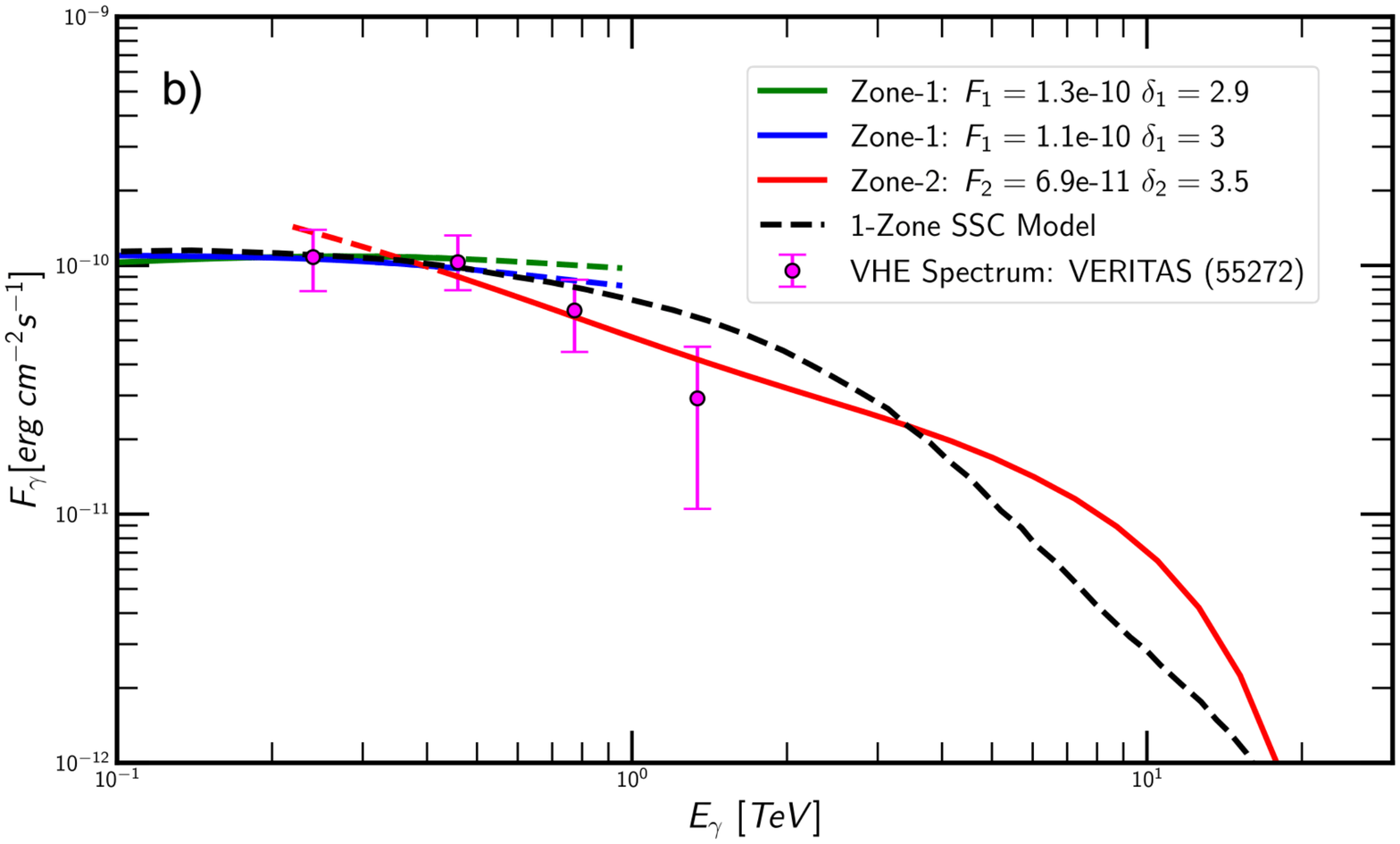}
\end{subfigure}%
\caption{The VHE spectra of MJD 55271 (a) and of MJD 55272 (b) observed by VERITAS telescopes are fitted with two-zone photohadronic model and shown along with their respective one-zone SSC fit for comparison.}
\label{fig:figure4}
\end{figure}
\begin{figure}
\begin{subfigure}
\centering
  \includegraphics[width=.5\linewidth]{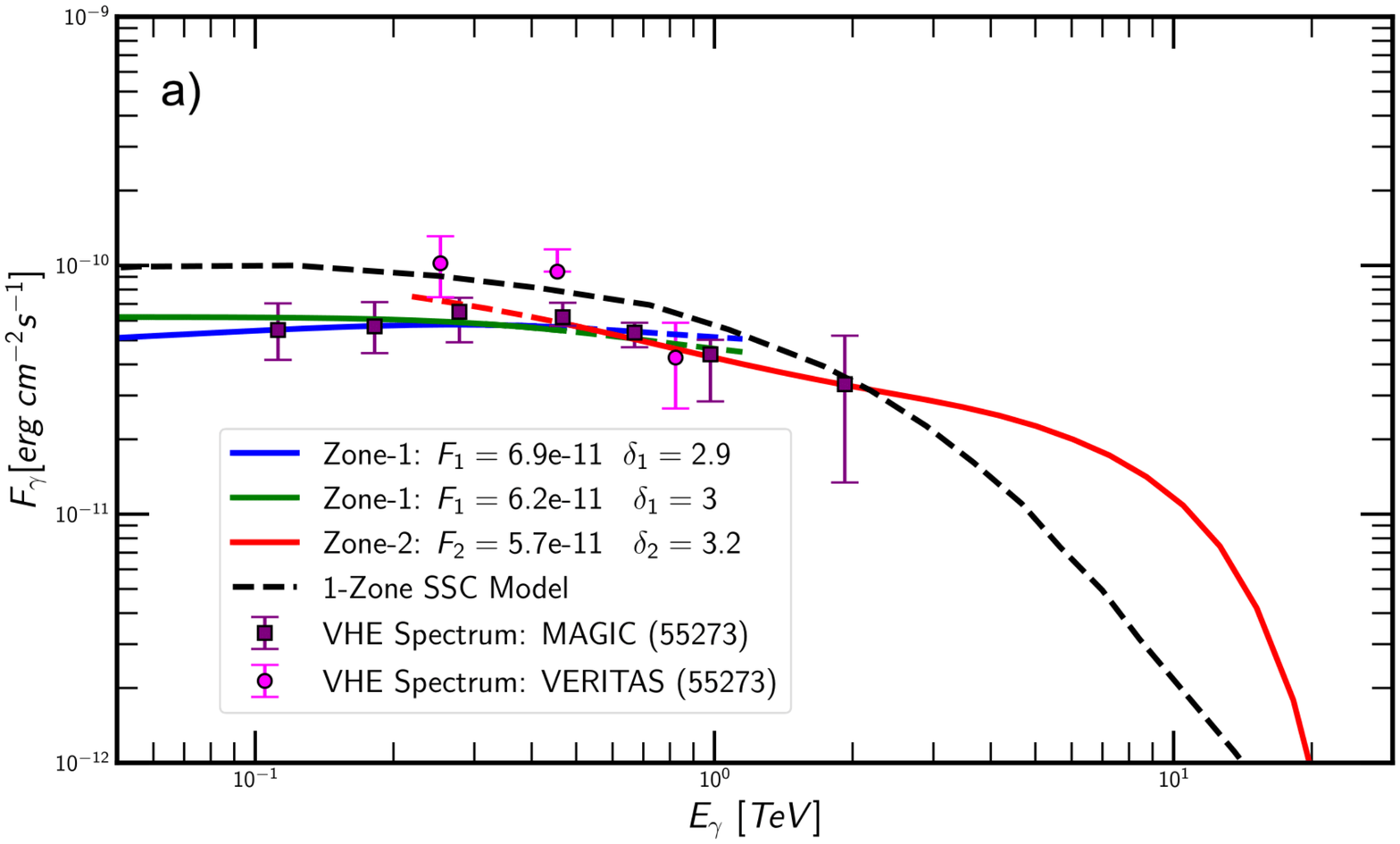}
\end{subfigure}
\begin{subfigure}
\centering
  \includegraphics[width=.5\linewidth]{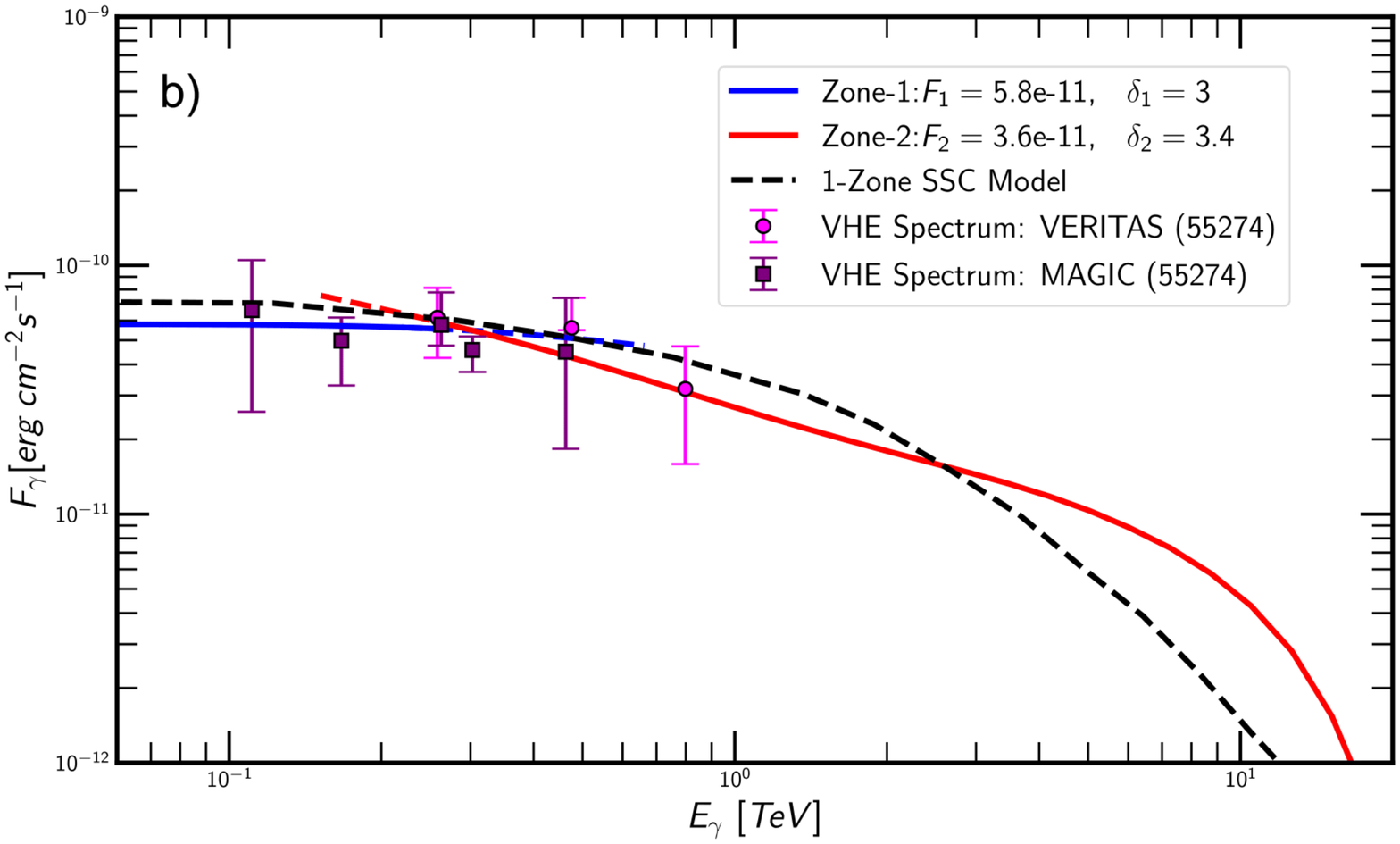}
\end{subfigure}%
\begin{subfigure}
\centering
  \includegraphics[width=.5\linewidth]{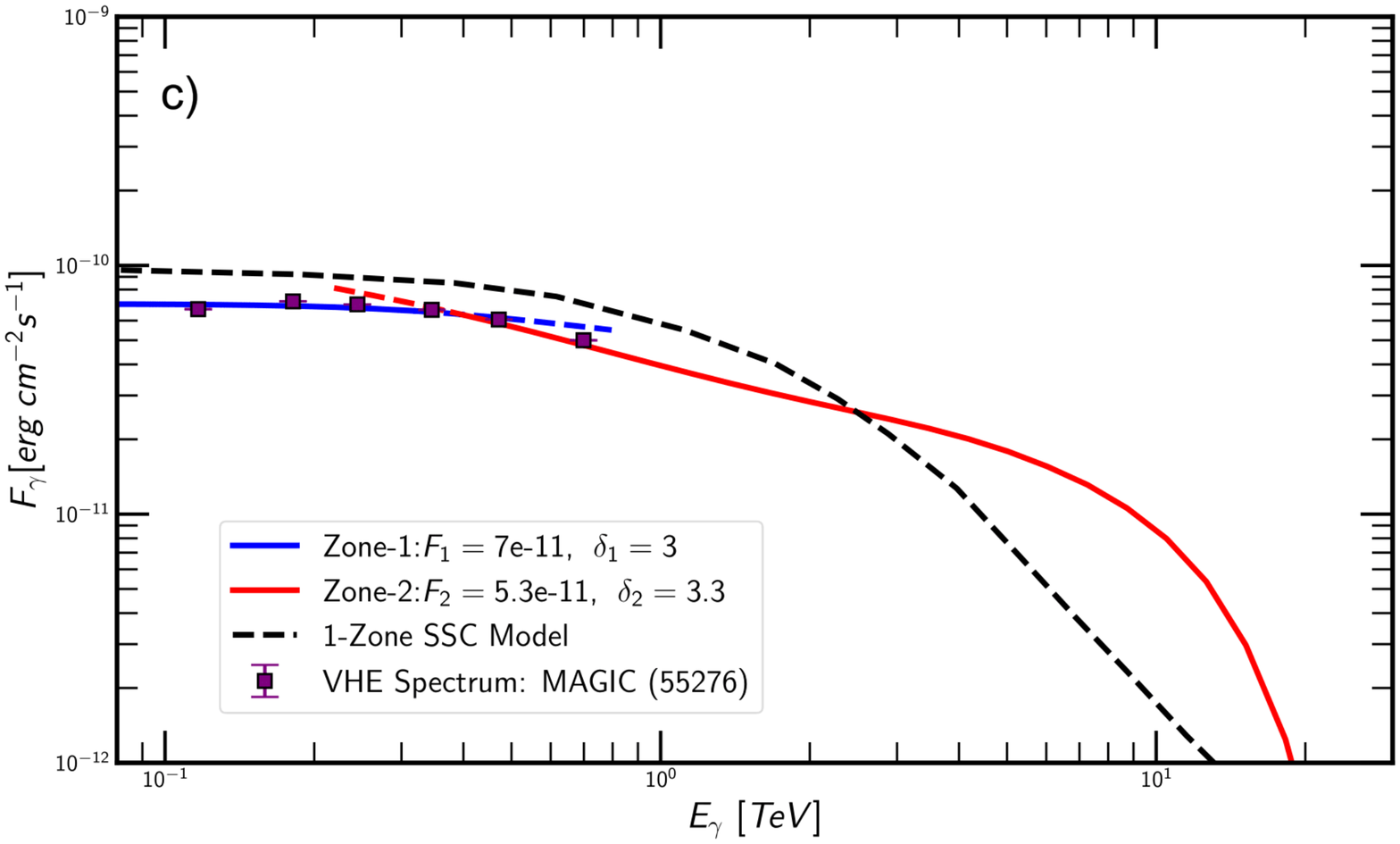}
\end{subfigure}%
\begin{subfigure}
\centering
  \includegraphics[width=.5\linewidth]{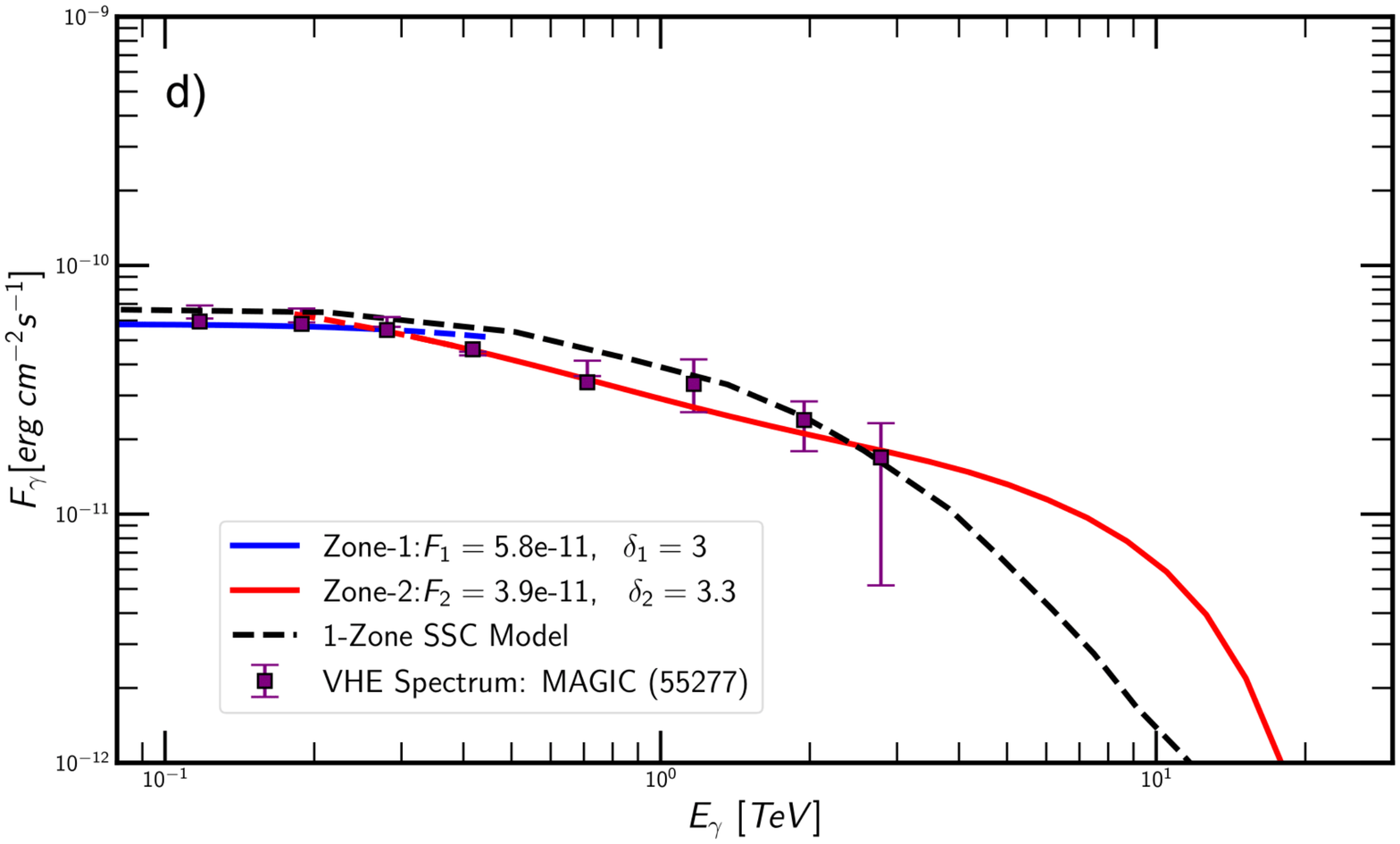}
\end{subfigure}
\caption{The VHE flaring data of MJD 55273 (a), MJD 55274 (b), MJD 55276 (c), and MJD 55277 (d) observed by the Cherenkov telescopes MAGIC and VERITAS are fitted with two-zone photohadronic model. For comparison, we have also plotted the one-zone SSC fits.
}
\label{fig:figure5}
\end{figure}

\section{Results}

During the MW campaign period of 2010, flaring in VHE gamma-rays was observed for 13 consecutive days from March 10 (MJD 55265) to March 22 (MJD55277). Initially the VHE flux was high ($\sim 2$ C.U.) and slowly the flux decreased to a typical value of $\sim 0.5$ C.U. towards the end of the flaring epoch.
In the VHE energy band, MAGIC, VERITAS and Whipple telescope systems observed the flaring events. The MAGIC telescopes made 11 observations and the exposure time for each period was about 10 to 80 min, amounting to a total of 4.7 h of good quality data. However, due to bad weather, data recorded on MJD 55272 and MJD 55275 were removed from the MW observations.

The VERITAS telescopes observed the flaring of Mrk 421 on MJD 55260, 55265, 55267-55274 with a 10 min run time each day. The observations were
performed at zenith angles $18^{\circ}-23^{\circ}$ to benefit from the lowest possible energy threshold of the events. Also, the Whipple 10 m telescope performed ten observations in ON/OFF and TRK (tracking) modes \citep{Pichel:2009gn}, lasting from one to six hours each on MJD 55267-55271 and MJD 55273-55277.

The most important aspect of these 13 days observation is that the synchrotron peak and the SSC peaks were displaced towards higher energies. Particularly, the synchrotron peak frequency was above $10^{17}$Hz and according to the classification scheme of the BL Lac objects, it belongs to EHBL category even though Mrk 421 is well known as an HBL. As discussed in the introduction, similar behavior has also been observed from Mrk 501 and 1ES 1959+650 which are well known HBLs, and there spectra are explained very well with the two-zone photohadronic scenario \citep{Sahu:2020tko,Sahu:2020kce}. Thus using the two-zone photohadronic model, the 13 days VHE flaring events observed by MAGIC and VERITAS are analysed here. For comparison, along with our results the one-zone SSC fits are also shown \citep{Aleksi__2015}. In this analysis, we ignore the statistical significance of a fit when only three or less observed data points are available.
 
The VHE spectrum of MJD 55265 (10 March 2010) observed by both MAGIC and VERITAS is in the energy range $0.1\, TeV\le E_{\gamma}\le 4.3 \, TeV$ and the synchrotron peak is at $\epsilon_{\gamma}\sim 4.6\times 10^{17}$ Hz which clearly shows its EHBL behavior. Using the two-zone photohadronic model, we fit the low-energy part of the spectrum (zone-1) (for $E^{intd}_{\gamma} \lesssim 1$ TeV) very well with $\delta_1=2.7$ and its corresponding flux normalization is $F_1=2.1\times 10^{-10}\ \mathrm{erg\,cm^{-2}\,s^{-1}}$, with a statistical significance of 98\% ($\chi^2=0.0008$). 
Also the best fit to the high energy part of the VHE spectrum (zone-2) is obtained for $F_2=2.0\times 10^{-10}\ \mathrm{erg\,cm^{-2}\,s^{-1}}$ and $\delta_2=3.2$, with a statistical significance of 96\% ($\chi^2=0.0366$). These are shown in Fig. \ref{fig:figure1}(a) along with the one-zone SSC model of \citep{Aleksi__2015} for comparison. 
It can be seen that our model fits better the data than the one-zone SSC model. The zone-1 and zone-2 have their respective seed SSC flux spectral indices $\beta_1=0.7$ and $\beta_2=1.2$ respectively. The VERITAS data behave differently in its highest energy when the flux has suddenly gone up and is non compatible with the two-zone model. However, previously it was shown that this can be fitted well with conventional photohadronic model with $F_0=2.1\times 10^{-10}\ \mathrm{erg\,cm^{-2}\,s^{-1}}$ and $\delta=2.6$ corresponding to a very high emission state \citep{Sahu:2019kfd}. As the MAGIC and VERITAS observation windows are seven hours apart, it is possible that intra-night variability might be responsible for this difference. 

On MJD 55266 the flaring event was observed by MAGIC telescopes only and the spectrum is in the energy range $0.1\, TeV\le E_{\gamma}\le 4.4$ TeV. The bow shape spectrum on this day (MJD 55266) can not be fitted by a single power-law. The zone-1 ($E^{intd}_{\gamma} \lesssim 1$ TeV) is fitted with $\delta_1=2.7$ and $F_1=2.5\times 10^{-10}\ \mathrm{erg\,cm^{-2}\,s^{-1}}$ having a statistical significance of 99\% ($\chi^2=0.0189$). Similarly, the best fit to zone-2 is obtained for $F_2=2.45\times 10^{-10}\ \mathrm{erg\,cm^{-2}\,s^{-1}}$ and $\delta_2=3.5$, with a statistical significance of 82\% ($\chi^2=0.0008$) and on this day the synchrotron peak is at $\epsilon_{\gamma}\sim 5.0\times 10^{17}$ Hz. Our fit and the one-zone SSC results are compared in Fig. \ref{fig:figure1}(b). It can be seen that the leptonic model is unable to fit the data in the low energy regime.

The only data available on MJD 55267 are from VERITAS and the spectrum is very well fitted with the two-zone photohadronic scenario with the transition energy $E^{intd}_{\gamma} \simeq 1$ TeV, which is shown in  Fig. \ref{fig:figure2}(a). The best fitted values for zone-1 are $F_1=2.5\times 10^{-10}\ \mathrm{erg\,cm^{-2}\,s^{-1}}$ and $\delta_1=2.7$ (statistical significance of 83\%, $\chi^2=0.0538$) while for zone-2 we have $F_2=2.5\times 10^{-10}\ \mathrm{erg\,cm^{-2}\,s^{-1}}$, and $\delta_2=3.3$. 
The SSC model of \cite{Aleksi__2015} predicts a VHE spectrum which falls significantly faster than the observed spectrum.

Both MAGIC and VERITAS observed the flaring event of MJD 55268 and the observed spectra are in the energy range $0.18\, TeV\le E_{\gamma}\le 2.1$ TeV. The MAGIC spectrum is well fitted in zone-1 with the spectral index $\delta_1=2.7$ and the normalization factor $F_1=1.3\times 10^{-10}\ \mathrm{erg\,cm^{-2}\,s^{-1}}$ 
and here $E^{intd}_{\gamma} \lesssim 0.8$ TeV . 
There are not many data points in zone-2, so we fitted this region with $\delta_2=3.4, 3.5$ 
and both values fit very well to the observed data. However, above 2 TeV there is a minor difference between these fits. Also our fit to the spectrum differs substantially from the one-zone SSC fit as shown in Fig. \ref{fig:figure2}(b).

The flaring event of MJD 55269 is very similar to the one observed on MJD 55268 and a good fit to the spectrum is achieved with $F_1=2.0\times 10^{-10}\ \mathrm{erg\,cm^{-2}\,s^{-1}}$  and $\delta_1=2.7$ for zone-1 (statistical significance of 96\%, $\chi^2=0.0790$), while $F_2=1.8\times 10^{-10}\ \mathrm{erg\,cm^{-2}\,s^{-1}}$, $\delta_2=3.2$ are for zone-2 
as shown in Fig. \ref{fig:figure3}(a).

On MJD 55270 the spectrum in the low energy regime i.e. zone-1 is flatter than previous days which implies that the flaring event is moving towards low emission state. 
However, because of less number of data points we fitted this region with $\delta_1=2.7, 2.8$ 
and both fit very well with very similar normalization constants ($F_1=1.32\times 10^{-10}\ \mathrm{erg\,cm^{-2}\,s^{-1}}$  and $1.16\times 10^{-10}\ \mathrm{erg\,cm^{-2}\,s^{-1}}$).
On the other hand, zone-2 is fitted well by $F_2=8.7\times 10^{-11}\ \mathrm{erg\,cm^{-2}\,s^{-1}}$, $\delta_2=3.3$ (statistical significance of 87\%, $\chi^2=0.0339$) as shown in Fig. \ref{fig:figure3}(b). One-zone SSC model also fits well to the spectrum. In the high energy regime, however, it falls faster than our fit. 

The flaring event of MJD 55271 was only observed by VERITAS in the energy range $0.25\, TeV\le E_{\gamma}\le 2.6$ TeV. Below the transition energy $E^{intd}_{\gamma} \lesssim 0.9$ TeV, the zone-1, we can fit well the spectrum with $F_1=(1.7-1.62)\times 10^{-10}\ \mathrm{erg\,cm^{-2}\,s^{-1}}$  and $\delta_1=(2.8-2.9)$. 
The zone-2 is fitted with $F_2\simeq 1.55\times 10^{-10}\ \mathrm{erg\,cm^{-2}\,s^{-1}}$ and  $\delta_2=3.3$. 
The one-zone SSC  model fit is also shown in Fig. \ref{fig:figure4}(a) to compare with our two-zone photohadronic model fit.  

For the observations from MJD 55272 to MJD 55277, with the exception MJD 55275 when there was no observation, the value of  $E^{intd}_{\gamma}$ is very low and in a narrow range of  $0.25\, TeV \lesssim E^{intd}_{\gamma}\lesssim 0.4$ TeV. Also the spectrum for each individual day is flat which is best fitted with either $\delta_1=2.9$ or $3.0$ and the statistical significance for these fits ranges from 85\% to 95\%. This value of $\delta_1$ signifies that the flaring of Mrk 421 has reached the low emission state and at the same time the seed photon flux in the SSC band has reached the maximum value of $\beta_1\simeq 1.0$. Also the maximum photon flux in zone-1 has substantially decreased compared to the first day of the observation. The zone-2 for these five days are also fitted very well with $3.2\lesssim \delta_2 \lesssim 3.5$ which correspond to the seed photons in the low energy tail region of the SSC band and their fluxes having spectral indices in the range $1.2\lesssim \beta_2\lesssim 1.5$. The flaring events of all these days along with the one-zone SSC fits are shown in Fig. \ref{fig:figure4}(b) and Figs. \ref{fig:figure5}(a), (b), (c) and (d) respectively.

Initially the VHE flaring was in high emission state and slowly it went down to low emission state by the end of the flaring period. This transition was accompanied by a gradual decrease in the transition energy $E^{intd}_{\gamma}$ from $\sim 1$ TeV to $\sim 0.25$ TeV and consequently zone-2 became wider and spread into zone-1 making the latter region narrower. We recall that the transition in the VHE spectrum took place due to the change in the seed SSC photon flux. So, 
physically what it means is that initially the photon flux in the tail region of the SSC band which is responsible for the $\Delta$-resonance production had two distinct energy dependencies as shown in Eq. (\ref{eq:sscflux}) (lower part with $\epsilon^{\beta_1}_{\gamma}$ and upper part with $\epsilon^{\beta_2}_{\gamma}$). Slowly the lower part drifted into the upper part by squeezing the former region into a narrow strip.

On MJD 55266, the maximum observed $\gamma$-ray energy was $E_{\gamma}\simeq 4.4$ TeV which is produced from the interaction of Fermi accelerated proton of energy $E_p\simeq 44$ TeV in the jet with the photons in the lower part of the SSC spectrum. In the two-zone SSC model of \cite{Aleksi__2015} for this day the SSC spectrum starts around $\epsilon_{\gamma}\sim 7\times 10^{21}$ Hz (Figure 8(b) of \cite{Aleksi__2015}). For this value of $\epsilon_{\gamma}$, the minimum value of the bulk Lorentz factor estimated in our model is $\Gamma\simeq 21$. Similar or smaller values for the bulk Lorentz factor are obtained for other days. On the other hand, in the one-zone SSC model (Figure 8(a) of \cite{Aleksi__2015}), the SSC spectrum starts above $2\times 10^{21}$ Hz and for this value of $\epsilon_{\gamma}$ we get minimum 
$\Gamma\simeq 35$ which is too high. 

\section{Discussion}

Mrk 421 is the nearest HBL and is the first extragalactic source to be observed in VHE. Since then it has undergone several episodes of multi-TeV flaring. Mrk 421 is also the first source observed in multiwavelength for exceptionally long period of time with dense monitoring. The VHE flaring epochs of this source during all its previous observations were consistently found to be in the HBL category which corresponds to the synchrotron peak in the frequency range $10^{15}-10^{16}$ Hz. However, during the 13 consecutive days observations undertaken during March 2010 the source was in the state of high activity with a changed spectral behavior and the synchrotron as well as the second peak shifted towards higher energies. The synchrotron peak shifted above $10^{17}$ Hz, implying that the blazar has undergone a shift from HBL to EHBL. Similar situations are observed in HBLs Mrk 501 and 1ES 1959+650, which were explained very well by the two-zone photohadronic scenario. For Mrk 421, we have used the same two-zone photohadronic model to interpret the 13 consecutive days of flaring in VHE and we find that the individual flaring events are explained very well by our model. This shows once more the success of two-zone photohadronic model in explaining the
extreme HBL nature of the VHE flaring events. We note that when the flaring in VHE was first observed on March 10, 2010 (MJD56265), the source was in high emission state and it continued to be in high emission state for about a week. Then it started degrading to low emission state when the flux was low and also the spectrum in zone-1 was almost flat. During the first week of the observation, the transition energy $E^{intd}_{\gamma}\simeq 1$ TeV when the flaring in the zone-1 was in high emission state. However, from MJD55272, the transition energy $E^{intd}_{\gamma}$ started decreasing and reached a minimum value of $0.25$ TeV on  MJD55274 and for rest of the days the zone-1 was in low emission state. The shift in the $E^{intd}_{\gamma}$ from $\sim 1$ TeV to $\lesssim 0.25$ TeV shows that towards the end of the VHE flaring, the zone-2 had dominated almost the whole spectrum, implying that the seed photon flux in SSC region has changed from a flatter behavior ($\beta_1\simeq 0.7$) to a steeper ($\beta_2\simeq 1.5$) one. This change in spectral behavior, in principle, can be detected by simultaneous observation of the SED in the low energy tail region of the SSC band and in the VHE band during the EHBL-like outburst of the source. We have also estimated the minimum bulk Lorentz factor, obtaining a reasonable value of $\Gamma\approx 21$. Further analysis of EHBL and EHBL-like transient behaviors will help solidify the evidence for the two-zone photohadronic origin of these phenomena.

We are thankful to Luis H. Castañeda Hernandez for many useful
discussions. The work of S.S. is
partially supported by DGAPA-UNAM (Mexico) Project No. IN103019. Partial support from CSU-Long Beach is gratefully acknowledged.

\bibliographystyle{aasjournal}
\bibliography{mrk421_ehbl_v3}{}
\end{document}